\documentclass[twocolumn,English,aps,pra,10pt,superscriptaddress,floatfix]{revtex4-2} 
\usepackage{times}
\usepackage{graphicx}
\usepackage{amssymb}
\usepackage{bm}
\usepackage{amssymb,amsfonts,amsmath,amsbsy,bm,t1enc,latexsym}
\usepackage{float}
\usepackage[colorlinks=true,citecolor=blue,linkcolor=magenta]{hyperref}
\usepackage[markup=blue, authormarkupposition=left]{changes}
\usepackage{url}
\usepackage[mode=text]{siunitx}
\DeclareSIUnit \dbc {dBc}
\DeclareSIUnit \dbm {dBm}
\DeclareSIQualifier\peak{p}
\usepackage{soul}
\usepackage{changes}
\usepackage{cancel}
\usepackage{times}
\usepackage{chemformula,xspace} 
\usepackage{braket}
\usepackage[capitalise]{cleveref} 

\begin{document}

\title{Chip-integrated single-mode coherent-squeezed light source using four-wave mixing in microresonators}

\author{Patrick Tritschler}
\email{patrick.tritschler@de.bosch.com}
\affiliation{Robert Bosch GmbH, Robert-Bosch-Campus 1, Renningen, 71272, Germany}
\affiliation{Institute for Micro Integration (IFM), University of Stuttgart, Allmandring 9b, Stuttgart, 70569, Germany}

\author{Torsten Ohms}%
\affiliation{Bosch Sensortec GmbH, Gerhard-Kindler Stra{\ss}e 9, Reutlingen, 72770, Germany}

\author{Christian Schweikert}%
\affiliation{Institute of Electrical and Optical Communications, Pfaffenwaldring 47, 70569 Stuttgart, Germany}

\author{Onur S\"ozen}%
\affiliation{Institute of Electrical and Optical Communications, Pfaffenwaldring 47, 70569 Stuttgart, Germany}

\author{Rouven H. Klenk}%
\affiliation{Institute of Electrical and Optical Communications, Pfaffenwaldring 47, 70569 Stuttgart, Germany}

\author{Simon Abdani}%
\affiliation{Institute of Electrical and Optical Communications, Pfaffenwaldring 47, 70569 Stuttgart, Germany}

\author{Wolfgang Vogel}%
\affiliation{Institute of Electrical and Optical Communications, Pfaffenwaldring 47, 70569 Stuttgart, Germany}

\author{Georg Rademacher}%
\affiliation{Institute of Electrical and Optical Communications, Pfaffenwaldring 47, 70569 Stuttgart, Germany}

\author{Andr\'{e} Zimmermann}%
\affiliation{University of Stuttgart, Institute for Micro Integration (IFM) and Hahn-Schickard, Allmandring 9b, Stuttgart, 70569, Germany}

\author{Peter Degenfeld-Schonburg}%
\email{peter.degenfeld-schonburg@de.bosch.com}
\affiliation{Robert Bosch GmbH, Robert-Bosch-Campus 1, Renningen, 71272, Germany}

\maketitle

\textbf{
Squeezed light constitutes a key resource for quantum optical technologies including quantum sensing, computing, communication and metrology. For many applications the generation of squeezed light typically requires at least two nonlinear optical stages involving careful phase and frequency matching to achieve the required mixing of squeezed and coherent light. In our work, we introduce an on-chip system that simplifies the generation of coherent-squeezed light, utilizing only a single squeezing stage. We achieve this by pumping a silicon nitride ($\mathrm{Si_3N_4}$) microring resonator to produce single-mode squeezed light through four-wave mixing at the same frequency as the pump mode, leveraging the inherent $\chi^{(3)}$-nonlinearity of the $\mathrm{Si_3N_4}$ resonator. Our on-chip system demonstrates a squeezing of -4.7 dB with a clear perspective towards -10 dB squeezing. We also provide a theoretical model that describes the straightforward yet robust generation of single-mode squeezing at the injection locking point of the ring resonator. In fact, we show that a design with a normal dispersion can be used for robust generation of bright squeezed light without the need for careful suppression of unwanted nonlinear processes. Overall, our findings highlight an approach which drastically simplifies the generation of coherent-squeezed light in photonic integrated circuits.
}

\section{Introduction}

Squeezed light is a well-established resource in sensor applications and quantum-metrology, where it enhances performance by reducing the noise \cite{PhysRevLett.116.061102, Vahlbruch_2010, SCHNABEL20171, Danilishin2012}. Its effectiveness has been demonstrated in various applications, including biosensors \cite{Li2020}, magnetometers \cite{PhysRevLett.127.193601}, general interferometers \cite{SCHNABEL20171, PhysRevLett.129.031101, tritschler2024optical, Tritschler:24}, and notably in the famous example of gravitational wave detection \cite{Vahlbruch_2010, PhysRevLett.116.061102, PhysRevLett.126.041102, Danilishin2012,  Goda2008, PhysRevLett.123.231108}. The enhancement in performance and the noise reduction is achieved through the squeezing process, which decreases the variance of one quadrature of the light, while increasing the variance of another quadrature \cite{2015_SqueezedLight, Andersen_2016, schnabel2023success, Schnabel2022, tritschler2024optical, SCHNABEL20171, Danilishin2012}. \\
Squeezed light is typically generated through two primary mechanisms: Three-wave mixing (TWM) \cite{Wu:87, SIZMANN1990138} or 
via Four-Wave Mixing (FWM) \cite{Bergman:91, PhysRevLett.77.3775, Garces2016, Hoff:15, PDDrummond_1980, Chen2023, Balybin2020}. In the TWM process, a single pump photon at a specific frequency $f_p$ is absorbed by a nonlinear medium with a significant second-order susceptibility $\chi^{(2)}$, leading to the emission of two photons. Similarly, in the FWM process, two pump photons at the frequency $f_p$ are absorbed by a nonlinear medium characterized by the third-order susceptibility $\chi^{(3)}$, resulting in the emission of two correlated photons. These emitted photons exhibit strong correlations, which leads to squeezing in the steady state of the output light, especially near the critical point of the classical dissipative phase transition. The TWM process consistently produces single-mode squeezed light in the case of degenerate down conversion, whereas the FWM process only achieves this when the emitted photons share the frequency with the pump $f_p$ as presented in this work. If the emitted photons differ in frequency, they are symmetrically spaced around $f_p$, resulting in the generation of two-mode squeezed light \cite{AGRAWAL2013397, Balybin2020}.\\
The TWM process is more commonly used as the second-order susceptibility $\chi^{(2)}$ is often several orders of magnitude larger than the third-order susceptibility $\chi^{(3)}$, which allows to generate squeezing at lower pump power. The generation of squeezed light through TWM has been successfully demonstrated in various contexts, including free-space optics \cite{SIZMANN1990138, PhysRevLett.59.278, PhysRevLett.59.2153, Goda2008, PhysRevLett.117.110801, PhysRevLett.123.231108, PhysRevLett.126.041102, PhysRevLett.129.031101, Zander_2023, chemosensors11010018} and in on-chip applications, particularly using periodically poled lithium niobate (PPLN) \cite{McKenna2022, Kashiwazaki2020, Kashiwazaki2021, Cheng:19, Mondain:19, Domeneguetti:23, Stokowski2023, Park2024}. The generation of single-mode squeezed light through FWM has been demonstrated in various macroscopic experiments, including those conducted in free space \cite{Villar2008, GALATOLA199195} and fiber optics \cite{GALATOLA199195, Kalinin2023, Fujiwara:09, Bergman:94, Bergman:91}. In this context, the squeezed light shares the same frequency as the pump, resulting in the formation of bright squeezed light \cite{PhysRevResearch.2.013371}. It is also possible to generate single-mode vacuum squeezed light by utilizing dual-pump FWM, that has been successfully demonstrated in chip-integrated ring resonators exploiting sophisticated measures for parasitic nonlinear process suppression \cite{Zhang2021, ulanov2025}. Two-mode vacuum-squeezed light via FWM has been successfully demonstrated in optical fibers \cite{PhysRevLett.77.3775, PhysRevLett.81.786, Sharping:01, Liu:16, Slavik2010} as well as in on-chip applications using microring resonators \cite{Yang2021, Vaidya_2020, Dutt:16, PhysRevApplied.3.044005}, even demonstrating two-mode squeezing across the entire frequency comb \cite{Liu:16, Yang2021, Guidry2022, Strekalov_2016, PhysRevLett.114.050501, PhysRevLett.108.083601, PhysRevLett.101.130501}. \\
Vacuum-squeezed light is more efficient for macroscopic applications. However, for sensor applications it is essential to mix this light with coherent light of the same frequency, either during or after the generation stage. This is because vacuum squeezed light alone does not offer a performance advantage compared to using the pump light directly and the mixing enables a performance improvement that can surpass the shot noise limit (SNL) \cite{schnabel2023success, PhysRevLett.116.061102, Vahlbruch_2010, SCHNABEL20171, Danilishin2012, tritschler2024optical}. This can be achieved by using two laser sources or by pumping a second nonlinear medium to generate another coherent laser, often through second harmonic generation (SHG) \cite{SCHNABEL20171, Danilishin2012, Zander_2023}. A key challenge in this approach is ensuring that both stages or sources are precisely controlled and frequency matched to facilitate efficient mixing and generate coherent-squeezed or rather bright squeezed light.  \\
In this work, we demonstrate the generation of single-mode coherent-squeezed light in a ring resonator via FWM, addressing the previously mentioned challenges by requiring only a single stage. This simplification enhances the overall control of the system and offers the advantages of a compact form factor and a high-quality factor resonator, resulting in a low threshold power due to significant field amplification \cite{10.1117/12.2656088}. The produced light features a strong coherent background, making it an ideal source for optical sensor applications. While the generation of single-mode coherent-squeezed light via FWM in microresonators has primarily been discussed theoretically \cite{Hoff:15}, we present experimental results from a silicon nitride ($\mathrm{Si_3N_4}$) microring resonator. We measure a squeezing of -1.219 dB, which corresponds to -4.7 dB of squeezing inside the chip at a pump power of 7.8 mW. We compare our measurements with theoretical predictions and demonstrate a strong agreement between the two. Finally, we show the simplicity and robustness of our system when driven at the injection locking point using a microring resonator in the normal dispersion regime. Our results pave the way for the development of a miniaturized, chip-integrated photonic solution for bright squeezed light sources, which can be utilized in quantum-enhanced sensor systems and emerging quantum technology applications.

\section{Results}
\begin{figure}
	\centering
	\includegraphics[width=0.49\textwidth]{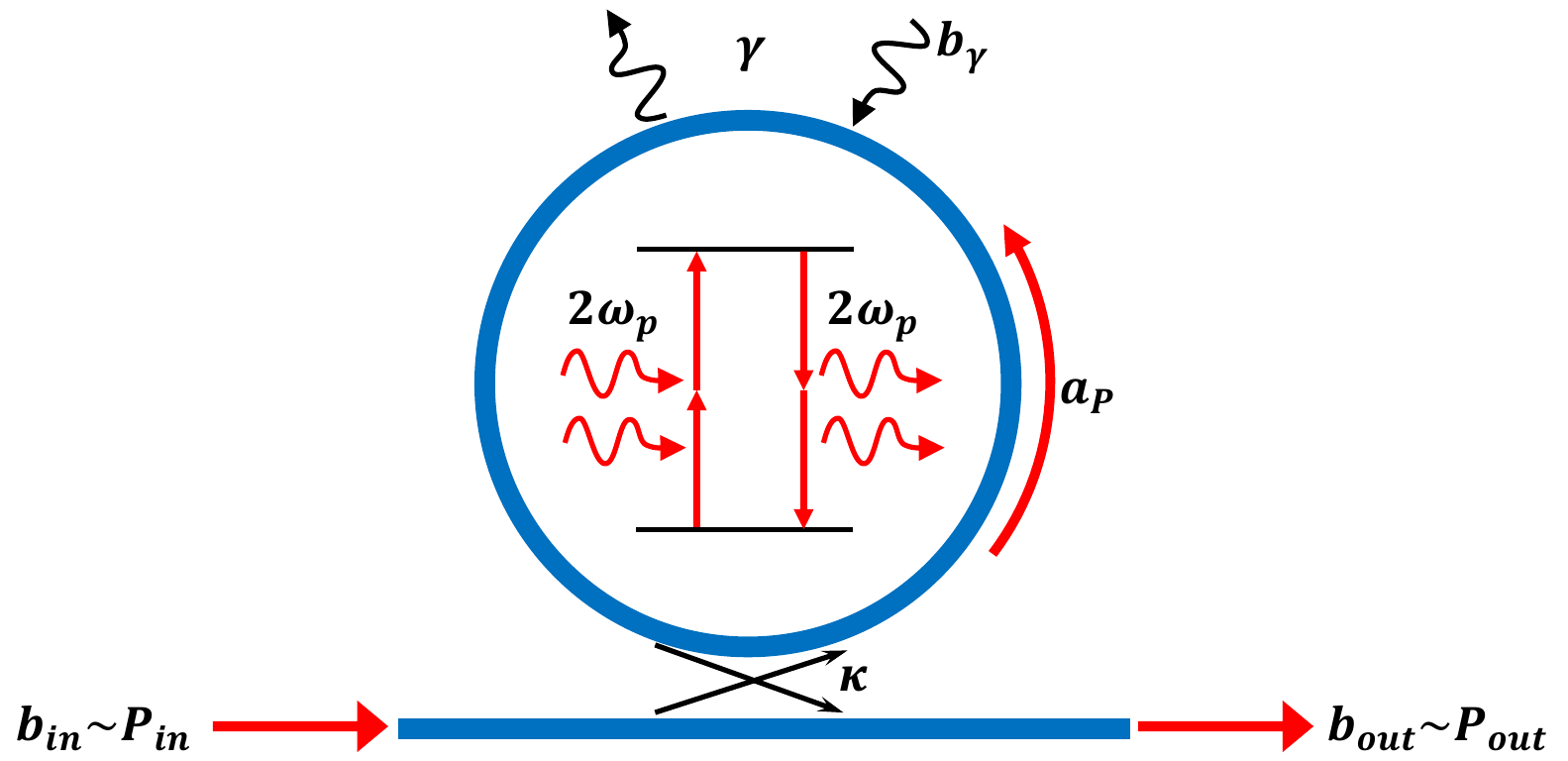}
	\caption{Model of the ring resonator. The ring is pumped by $P_\mathrm{in}$ over $\kappa$ which excites the cavity mode $a_p$ and leads to the FWM process inside the ring resonator. Losses appear through $\kappa$ and $\gamma$ which cause a mixing of $a_p$ with the vacuum modes $b_\gamma$ and $b_\mathrm{in}$. The field $b_\mathrm{out}$ describes the output power of the light.}\label{fig:ring_model_main}
\end{figure}
In the following sections, we will first analyze the theoretical results to assess the achievable squeezing performance based on our $\mathrm{Si_3N_4}$ material platform and the designed geometry. In our model equations, however, we will determine the parameters from experimental characterization techniques rather than relying on numerical simulations. Subsequently, the squeezing measurements are presented and compared with the theoretical predictions, thereby validating the derived model.

\subsection{Single-Mode squeezing via Four-wave mixing in microring resonators}\label{sec:single-mode-squeezing-via-four-wave-mixing-in-microring-resonators}
\begin{figure*}
	\centering
	\includegraphics[width=1\textwidth]{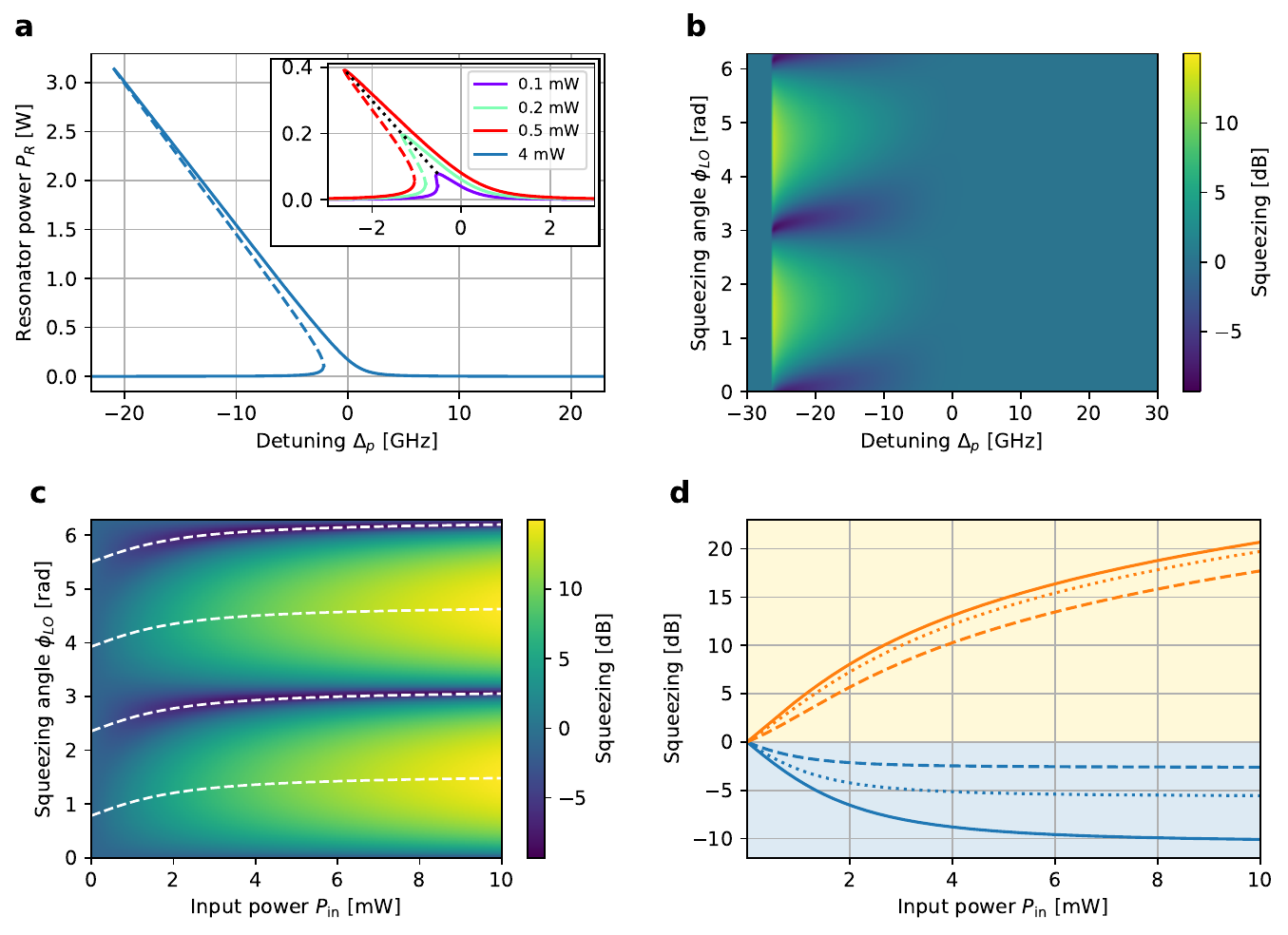}
	\caption{Optical power and output squeezing spectrum generated via FWM in the ring resonator. All values are determined for $\kappa=500$ MHz, $\gamma=50$ MHz, $g_\mathrm{opt}=1.5$ Hz, $g_\mathrm{th}=100$ Hz, $R=211$ {\textmu}m and $n_\mathrm{eff}=1.9$. (a) Optical power inside the resonator $P_R$ in dependency of the Detuning $\Delta_p = \omega_p - \omega_R$ at different input powers denoted by the colors. The injection locking points are marked by the black dotted line. (b) Squeezing spectrum in dependency of the detuning $\Delta_p$ and the squeezing angle $\phi_{\mathrm{LO}}$ at an input power of $P_\mathrm{in}=4$ mW. (c) Squeezing spectrum at the injection locking point in dependency of the input power $P_\mathrm{in}$ and the squeezing angle $\phi_{\mathrm{LO}}$. The optimal phase for the squeezing $\phi_{\mathrm{LO, opt}}$ and anti-squeezing $\phi_{\mathrm{LO, opt}}+\pi/2$ is marked by the white dashed lines. (d) Squeezing (blue lines) and anti-squeezing (orange lines) at various efficiencies with $\eta=1$ (solid line), $\eta=0.8$ (dotted line) and $\eta=0.5$ (dashed line) in dependency of the input power $P_\mathrm{in}$ at the optimal phase $\phi_{\mathrm{LO, opt}}$.}\label{fig:calculated_squeezing3}
\end{figure*}
\begin{figure*}
	\centering
	\includegraphics[width=1\textwidth]{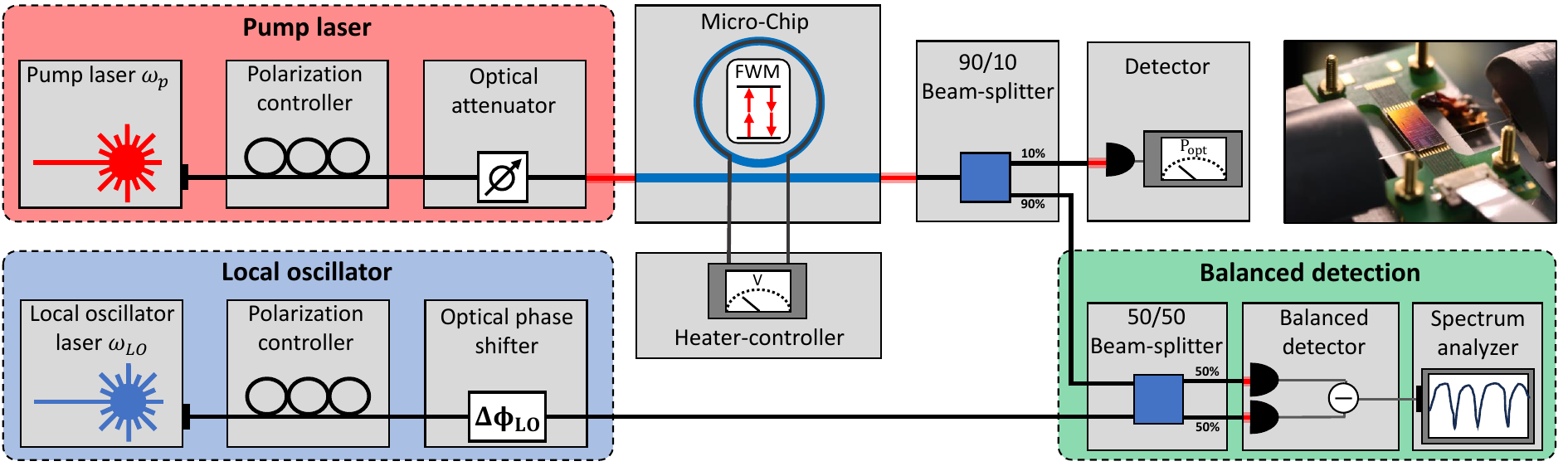}\\
	\caption{Overview of the experimental setup. The light from a pump laser is sent through a polarization controller and an optical optical attenuator inside of the micro-chip. The resonance condition of the ring can be controlled by a metal heater above which is controlled using a power supply. From the outcoupled light, 10\% are observed at a detector to verify the detuning $\Delta_\mathrm{cl}$ while 90\% are mixed at a 50/50 beam splitter with a LO for the balanced detection. The LO consists of a tuneable laser source which is sent through a polarization controller and an optical phase shifter which is controlled by a power supply. The mixed light is measured using a balanced detector and observed at a spectrum analyzer. }\label{fig:experimental_setup}
\end{figure*}
In its most basic form, a ring resonator consists of a ring-shaped optical waveguide connected to a straight waveguide, as shown in Figure \ref{fig:ring_model_main}. The resonator is pumped by an input mode $b_\mathrm{in} = \beta_\mathrm{in} + \delta b_\mathrm{in}$ with the coherent laser pump $\beta_\mathrm{in}=\sqrt{P_\mathrm{in}/\hbar\omega_{p}}$, where $P_\mathrm{in}$ represents the optical power and $\omega_{p}$ denotes the angular laser frequency to excite the resonator mode $a_p$. As commonly done, we separate the bosonic light field $a_p = \alpha_p + \delta a_p$ into its coherent part $\alpha_p$ and its quantum fluctuations $\delta a_p$ and model the dynamics within rotating wave approximation by the linearized quantum Langevin equations describing on one hand the classical dynamics of the coherent part with
\begin{eqnarray}
\frac{d }{dt}\alpha_{p} = i \Delta_\mathrm{cl} \alpha_{p} - \frac{\Gamma}{2} \alpha_p + \sqrt{\kappa} \beta_\mathrm{in} + \sqrt{\gamma} \beta_{\gamma}, \label{equ:quantumLangevin_SpmClassic_linearized_main}
\end{eqnarray}
and on the other hand the quantum dynamics of the fluctuations with
\begin{equation}
\frac{d }{dt} \delta a_{p} = i\Delta_f  \delta a_p + ig_{\mathrm{opt}}  \alpha_{p}^2 \delta a_p^\dagger -  \frac{\Gamma}{2}\delta a_p + \sqrt{\kappa} \delta b_\mathrm{in} + \sqrt{\gamma}\delta b_{\gamma}, \label{equ:quantumLangevin_SpmSqueezing_linearized_main}
\end{equation}
as we discuss in more detail in the supplement materials. In these equations, $\beta_\mathrm{in}$, $\beta_\gamma$ and $\alpha_p$ represent the classical mode amplitudes, while $\delta a_p$, $\delta b_\mathrm{in}$ and $\delta b_\gamma$ correspond to the quantum fluctuations. The resonator modes experience losses characterized by the coupling rate $\kappa$ and the loss rate $\gamma$, which together define the total loss rate $\Gamma=\kappa+\gamma$. The vacuum mode $b_\gamma$ with $\beta_{\gamma}=0$ enters the ring through the loss rate $\gamma$, while the input mode $b_\mathrm{in}$ couples via the coupling rate $\kappa$. The detuning for the classical mode is given by $\Delta_\mathrm{cl} =  \omega_{p}-\omega_{R} + |\alpha_p|^2 (g_\mathrm{opt} + g_\mathrm{th})$, while for the fluctuations, it is expressed as $\Delta_f = \Delta_\mathrm{cl} + g_{\mathrm{opt}} |\alpha_{p}|^2$. Here, $\omega_{R}$ denotes the cold cavity resonance frequency, and $g_{\mathrm{opt}}$ and $g_\mathrm{th}$ represent the optical and thermal gain, respectively \cite{tritschler2024nonlinearopticalbistabilitymicroring}. \\
The squeezing occurs through the Kerr effect, represented by the term $i g_\mathrm{opt}\alpha_{p}^2 \delta a_p^\dagger$ in equation \ref{equ:quantumLangevin_SpmSqueezing_linearized_main}. Therefore, a high classical amplitude $\alpha_{p}^2$ within the ring is essential for significant squeezed light generation. To optimize this process, it is most efficient to keep the laser frequency locked to the shifted resonance frequency, accounting for the nonlinear Kerr effect $\sim g_\mathrm{opt}$ and thermal self-heating $\sim g_\mathrm{th}$, ideally achieving $\Delta_\mathrm{cl} = 0$. With a proper control loop, this so-called injection locking point can be kept stable over external influences on the ring resonator, such as temperature variations. \\
In a first step, it is important to note that $\Delta_f$ is always larger than $\Delta_\mathrm{cl}$ by a factor of $g_\mathrm{opt}|\alpha_{p}|^2$, indicating that the fluctuations are consistently detuned. Thus, the detuning in our model is different from the detuning present in cases with TWM or FWM where it is engineered to be close to zero for optimal squeezing generation. \\
We solve equations \ref{equ:quantumLangevin_SpmClassic_linearized_main} and \ref{equ:quantumLangevin_SpmSqueezing_linearized_main} for the squeezing, which leads to the variance $\Delta X_Q$ of the output quadrature operator $X_Q(\phi_{LO}) = 1/\sqrt{2} [\delta b_\mathrm{out} e^{i\phi_{LO}} + \delta b_\mathrm{out}^\dagger e^{-i\phi_{LO}}]$ that depends on the output fluctuations $\delta b_\mathrm{out}$ and the squeezing angle $\phi_{LO}$. The detailed derivations are shown in the supplement materials and the results are shown in Figure \ref{fig:calculated_squeezing3}/b, together with the optical power inside of the resonator in Figure \ref{fig:calculated_squeezing3}/a as a function of the cold cavity detuning defined by $\Delta_P = \omega_p - \omega_R$. The results reveal a hysteresis in the classical solution known as optical bistability or self-phase modulation \cite{tritschler2024nonlinearopticalbistabilitymicroring}, which consequently also appears in the squeezing spectrum. Notably as seen in Figure \ref{fig:calculated_squeezing3}/b, the squeezing is broadband, indicating that squeezed light generation remains feasible despite the detuning $\Delta_f$ for the resonator quantum fluctuation field introduced by $g_\mathrm{opt}|\alpha_{p}|^2$. Therefore, it is not essential to operate the ring resonator at the injection locking point $\Delta_\mathrm{cl}=0$ and a certain degree of detuning is acceptable to generate significant squeezing. However, it becomes evident that optimal squeezing appears at the injection locking point. \\
Furthermore, Figure \ref{fig:calculated_squeezing3}/b shows that the squeezing angle $\phi_{LO}$ at which the optimal squeezing is achieved, varies with the detuning. Typically, the optimal squeezing occurs at a fixed squeezing angle of $\phi_{LO}=\pi/2$ \cite{Walls2008}. However, due to the additional detuning in our system, the optimal squeezing angle becomes dependent on $\Delta_f$ and, consequently, on the input power. Remarkably, at the injection locking point we are able to present analytical results. Introducing the threshold power $P_\mathrm{th} = \Gamma^3\hbar\omega_p/(8g_\mathrm{opt}\kappa)$, which describes the power needed for the classical FWM to generate a frequency comb \cite{tritschler2024optical, Herr2012}, we find the optimal squeezing angle at $\phi_{\mathrm{LO, opt}} = (1/2)\tan^{-1}(- P_\mathrm{th}/(2P_\mathrm{in}))$. Finally, we determine the squeezing $V_\mathrm{s}/V_\mathrm{vac}=2\Delta X_Q(\phi_{\mathrm{LO, opt}})$ and anti-squeezing $V_\mathrm{as}/V_\mathrm{vac}=2\Delta P_Q(\phi_{\mathrm{LO, opt}})=2\Delta X_Q(\phi_{\mathrm{LO, opt}}+\pi/2)$, which are given by 
\begin{eqnarray}
\begin{aligned}
\frac{\langle V_\mathrm{s} \rangle }{\langle V_\mathrm{vac}\rangle} & =  1 - \frac{8\eta\kappa}{\Gamma}\left(\frac{P_\mathrm{in}}{P_\mathrm{th}}\right)^2 \left( \sqrt{1+\left(\frac{P_\mathrm{th}}{2P_\mathrm{in}}\right)^2} - 1 \right)&\\ & \xrightarrow[]{P_\mathrm{in}\rightarrow \infty} 1 - \frac{\eta\kappa}{\Gamma},&\label{equ:squeezing}
\end{aligned}
\end{eqnarray}
\begin{eqnarray}
\begin{aligned}
\frac{\langle V_\mathrm{as} \rangle }{\langle V_\mathrm{vac}\rangle} & =  1 + \frac{8\eta\kappa}{\Gamma}\left(\frac{P_\mathrm{in}}{P_\mathrm{th}}\right)^2 \left( \sqrt{1+\left(\frac{P_\mathrm{th}}{2P_\mathrm{in}}\right)^2} + 1 \right) & \\ &\xrightarrow[]{P_\mathrm{in}\rightarrow \infty} \infty. &\label{equ:antisqueezing}
\end{aligned}
\end{eqnarray}
Both equations are normalized by the variance of the vacuum mode $\langle V_\mathrm{vac} \rangle$ and include the efficiency $\eta$, which describes the losses during the out-coupling from the ring up to the detection system. All the detailed calculations leading to this main result can be found in the supplement materials. The results for the squeezing are presented in Figure \ref{fig:calculated_squeezing3}/c, showing the behavior at the injection locking point for varying input power $P_\mathrm{in}$ as a function of $\phi_{\mathrm{LO}}$. Additionally, Figure \ref{fig:calculated_squeezing3}/d illustrates the results at the optimal local oscillator phase $\phi_{\mathrm{LO, opt}}$. It is clear that the squeezing decreases and the anti-squeezing increases with $P_\mathrm{in}$. For a significant squeezing it is desired to operate at input powers close or beyond the threshold power with $P_\mathrm{in} > P_\mathrm{th}$, and thus, it is recommended to utilize a ring design with normal dispersion to ensure that the FWM process, along with other cavity modes, does not interfere with the squeezing generation. In fact, the single-mode FWM process will also never reach its threshold in an asymptotical sense due to the finite detuning $\Delta_f$ for all input powers. The distance $x$ to the critical point, with $x=1$ marking the critical point, of the single-mode FWM process is simply given by $x^2 = 1/(1+(P_\mathrm{th}/P_\mathrm{in})^2)$ and thus stays always below one. The resonator in the normal dispersion regime can therefore be driven into the large power regime where the squeezing varies minimal but the brightness of the output light still scales linearly with the laser power making our system highly robust and easy to control.\\
To achieve optimal squeezing levels, it is essential for the resonator design to have a high coupling rate that significantly exceeds the optical losses with $\kappa \gg \gamma$. Additionally, a high efficiency approaching $\eta \rightarrow 1$ is desired, as illustrated in Figure \ref{fig:calculated_squeezing3}/D. These requirements align with findings from other squeezed light sources \cite{SCHNABEL20171}. Also note that the minimal uncertainty state is reached in the limit $\Gamma \rightarrow \kappa$ and $\eta \rightarrow 1$ for all input powers $P_\mathrm{in}$.

\subsection{Single-Mode squeezed light generation}
\begin{figure*}
	\centering
	\includegraphics[width=1\textwidth]{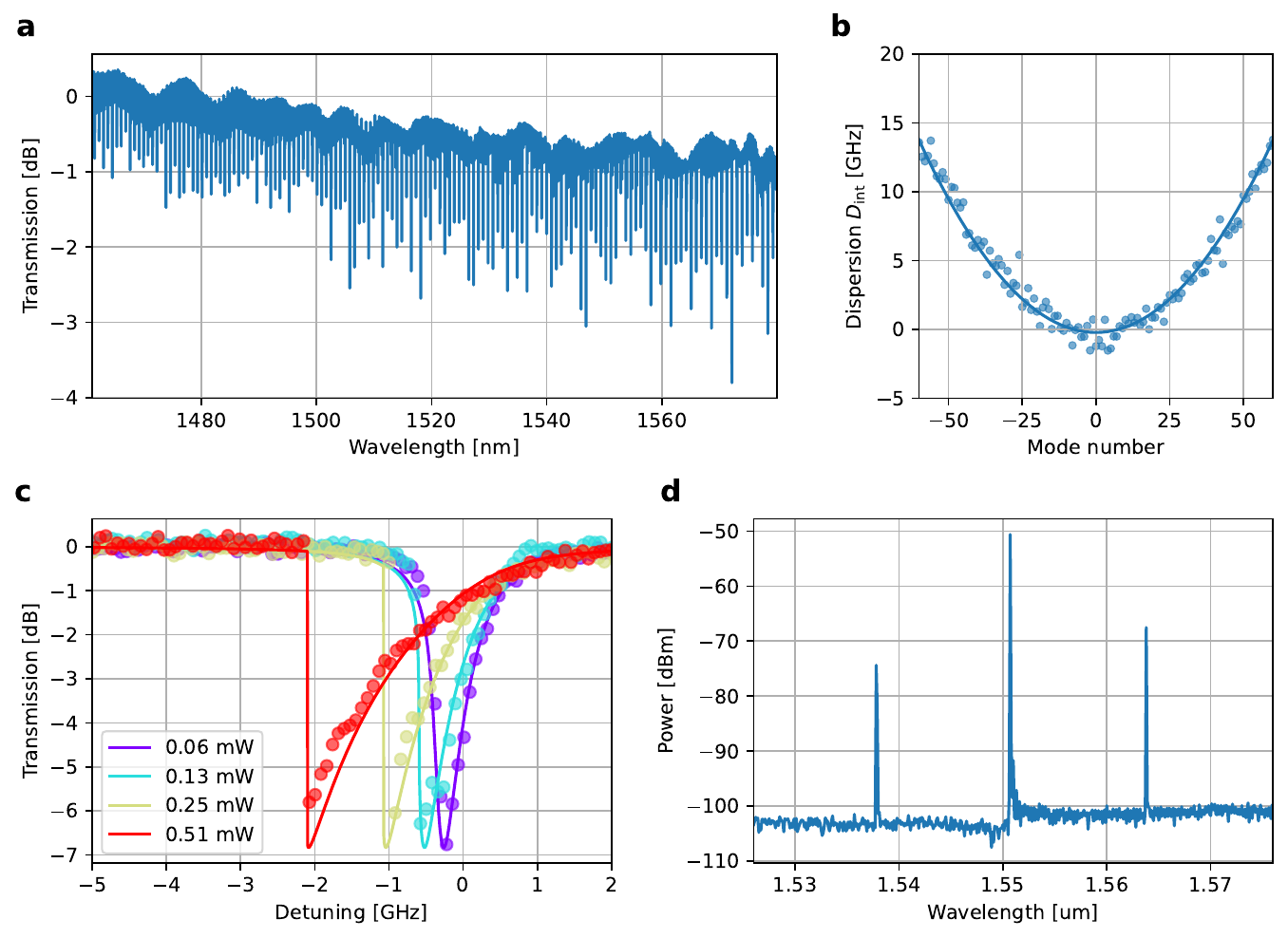}
	\caption{Characterization measurements of the ring resonator. (a) Measured transmission spectrum of the ring resonator. (b)  Measured (fitted) dispersion as dots (line) over the mode numbers, which shows the anomalous dispersion with $D_2 = 7.76$ MHz. (c) Transmission measurements at various input power to extract the resonator parameters $\kappa=515$ MHz, $\gamma=192$ MHz and $g_\mathrm{th}=127$ Hz. (d) Measured frequency comb using an optical spectrum analyzer. The comp appears at the threshold power of $P_\mathrm{th}\approx7.89$ mW, which leads to the optical gain $g_\mathrm{opt}=1.4$ Hz.}\label{fig:chip_charakterization}
\end{figure*} 
\begin{figure*}
	\centering
	\includegraphics[width=0.45\textwidth]{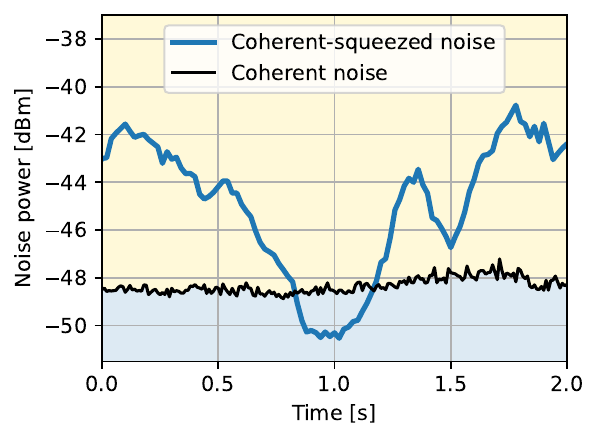}
	\includegraphics[width=0.43\textwidth]{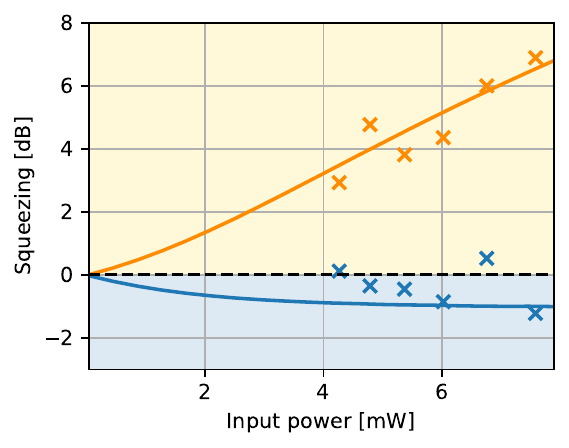}
	\caption{Results of the squeezing measurement using homodyne measurement. The black line represents the reference measurement of the coherent noise while the measurement in the blue (orange) field corresponds to squeezed (antisqueezed) measurements. \textbf{Left:} Zero-span measurement in the spectrum analyzer. The noise power is measured over time which corresponds to a sweep of the LO phase $\phi_{\mathrm{LO}}$. The minimum value of the measurement curve corresponds to the squeezed quadrature whereas the maximum defines the anti-squeezed noise. The difference between the squeezing and anti-squeezing measurements with respect to the coherent noise level leads to the squeezing $V_\mathrm{s} / V_\mathrm{vac}$ and anti-squeezing $V_\mathrm{as} / V_\mathrm{vac}$ in dB as displayed in the right plot. \textbf{Right:} Results over different input power with the measurements as crosses and the theoretical results as lines for $V_\mathrm{s} / V_\mathrm{vac}$ in orange and $V_\mathrm{as} / V_\mathrm{vac}$ in blue color.}\label{fig:measurement_results}
\end{figure*}
We validate our concept and the theoretical results using a $\mathrm{Si_3N_4}$ ring resonator with a width of 1.6 {\textmu}m, a height of 800 nm and a radius of 211 {\textmu}m, connected to a straight waveguide via a gap distance of 0.52 {\textmu}m. The motivation behind this design is that the selected geometry exhibits anomalous dispersion, which enables the generation of a frequency comb. This is crucial for experimentally characterizing $g_\mathrm{opt}$ through the threshold power $P_\mathrm{th}$ required for the FWM process. To achieve significant squeezing, it is essential that $\kappa$ is larger than $\gamma$, as discussed in equation \ref{equ:squeezing}. Therefore, we choose a small gap distance in combination with a large waveguide width to create an overcoupled resonator with low losses. However, for practical purposes $\kappa$ can not be chosen arbitrarily large since the threshold power $P_\mathrm{th}$ scales quadratically with $\kappa$. \\
The experimental setup used to characterize the micro-chip is illustrated in Figure \ref{fig:experimental_setup}. It comprises different stages for pumping the chip and to perform the measurements. In the pump laser stage, a laser with a wavelength of $\lambda_{p}=1550$ nm and an optical power of 100 mW is directed through a polarization controller and an optical attenuator before entering the micro-chip. Lensed fibers are used to enhance the coupling efficiency. Inside the chip, light couples from the waveguide to the ring resonator, where single-mode squeezed light is generated via FWM. To control the resonance frequency of the resonator $\omega_{R}$, a metal heater is positioned above the ring and connected to a heater-controller. After coupling out the light, 10 \% of the power is directed to a detector to monitor the detuning $\Delta_\mathrm{cl}$, facilitating squeezed light generation near the injection locking point. The remaining 90 \% of the light is mixed at a 50/50 beam splitter with a local oscillator (LO), which consists of a laser source operating at 1 mW and the frequency $\omega_{LO}$. This light passes through a polarization controller and an optical phase shifter before reaching the 50/50 beam splitter. The analyzed squeezing angle can be adjusted by varying the phase $\phi_{LO}$ applied by the phase shifter. The mixed light is then detected using a balanced detector and the resulting photocurrent signal is analyzed with a spectrum analyzer. The frequency $\omega_{LO}$ is tuned close to $\omega_{p}$ to perform a heterodyne-measurement \cite{wiseman_milburn_2009}. \\
In addition to the squeezing measurements, the ring resonator is characterized to determine the relevant physical parameters of the ring. This involves measuring the transmission spectrum and generating a frequency comb, both of which are presented in Figure \ref{fig:chip_charakterization}. The transmission measurement was conducted at various input powers, with a gradual reduction of the pump frequency $\omega_{p}$. The optical bistability becomes clearly visible at higher power levels. By fitting the spectrum using the classical equation of motion given by equation \ref{equ:quantumLangevin_SpmClassic_linearized_main}, we determine $\kappa=515$ MHz and $\gamma=192$ MHz, which corresponds to a quality factor of $Q=1.7\cdot10^6$. This indicates that the resonator is overcoupled, as required for significant squeezed light generation. Using the transmission curves obtained at different input powers, we also fit the thermal gain to $g_\mathrm{th}=127$ Hz. To determine the optical nonlinearity, we generate a frequency comb, replacing the balanced detection with an optical spectrum analyzer. Using the transmission measurements presented in Figure \ref{fig:chip_charakterization}/a, we determine the dispersion characteristics shown in Figure \ref{fig:chip_charakterization}/b, confirming the anomalous dispersion. The first generated side modes appear at the threshold power of $P_\mathrm{th}\approx 7.89$ mW, leading to an optical gain of $g_\mathrm{opt}=1.4$ Hz. \\
Finally, single-mode squeezing measurements are performed at various input powers $P_\mathrm{in}$, which are adjusted using the optical attenuator. The signal is captured using the spectrum analyzer in the zero-span mode at a frequency of 100 MHz. The phase of the LO $\phi_{LO}$ is continuously varied and a reference measurement is taken for each power setting as the level of coherent noise varies together with $P_\mathrm{in}$. The results are presented in Figure \ref{fig:measurement_results}, with a raw time measurement displayed on the left. It is clearly visible that the noise of the generated squeezed light varies over time and thus, over changes in $\phi_{LO}$. The values within the blue shaded area indicate squeezing, while those in the orange shaded area represent anti-squeezing. Although we encountered some phase noise within the system which contributed additional noise to the measurements, the squeezing and anti-squeezing effects remain clearly visible. Since the reference measurements also exhibit fluctuations, it is necessary to remove this systematic error from the squeezing measurements.\\
The evaluated squeezing results at various input powers are presented as crosses in the right plot of Figure \ref{fig:measurement_results}, alongside the theoretical predictions derived from the experimentally characterized parameter values as lines. At an input power of $P_\mathrm{in}=7.59$ mW, thus slightly below the threshold power $P_\mathrm{th}\approx7.89$ mW, we measure a squeezing of -1.219 dB and anti-squeezing of 6.89 dB. By determining the efficiency of our system to $\eta\approx 29.1$ \%, we estimate the squeezing within the resonator to be approximately -4.7 dB. As the input power decreases, both squeezing and anti-squeezing levels are also reduced. This, finally, indicates that we have successfully generated a squeezed state of light through FWM within the resonator.

\section{Discussion}\label{sec12}

The measurements show that a squeezing of -4.7 dB is achieved within the microchip when the ring resonator is pumped with an input power of $P_\mathrm{in}=7.59$ mW. This result aligns well with theoretical predictions and is possible due to the high Q-factor and the overcoupled design of the resonator with $\kappa > \gamma$. Note that our model does not rely on any fit parameters and the main result for the variances given by simple analytic expressions in equations \ref{equ:squeezing} and \ref{equ:antisqueezing} align remarkably well with the measurements. The generated fluctuations are mixed with the coherent background in the waveguide, resulting in the production of coherent squeezed light. In the normal dispersion regime, our system is stable and the interference between various cavity modes is suppressed. A further significant advantage of the proposed system is that it requires only a single pump source and one resonator for coherent squeezed light generation. This simplifies the required system control, reduces production costs and enhances the efficiency, while also increasing the usability through various integration possibilities. Consequently, this system can be applied in numerous optical sensor applications, improving the performance while minimizing the size and complexity of the squeezing source. \\
However, it is crucial to emphasize that this approach is only effective if the generated squeezed light experiences minimal losses up to the detection stage. In our setup, significant squeezing is lost during the out-coupling of the light, which limits the overall performance of the source. For enhanced applications, we recommend either integrating a complete sensor system within the chip or optimizing the light coupling. When further combined with a compact integrated pump laser source, this approach enables the emission of high-power single-mode coherent squeezed light. Overall, the proposed system has the potential to significantly expand the application of quantum-enhanced sensor systems driven by a squeezed light source due to its small form factor, low pump power requirements, and the overall simplicity.
 
\medskip

\section{Methods}\label{sec11}

\subsection*{Experimental setup}

\begin{figure}[h]
	\centering
	\includegraphics[width=0.5\textwidth]{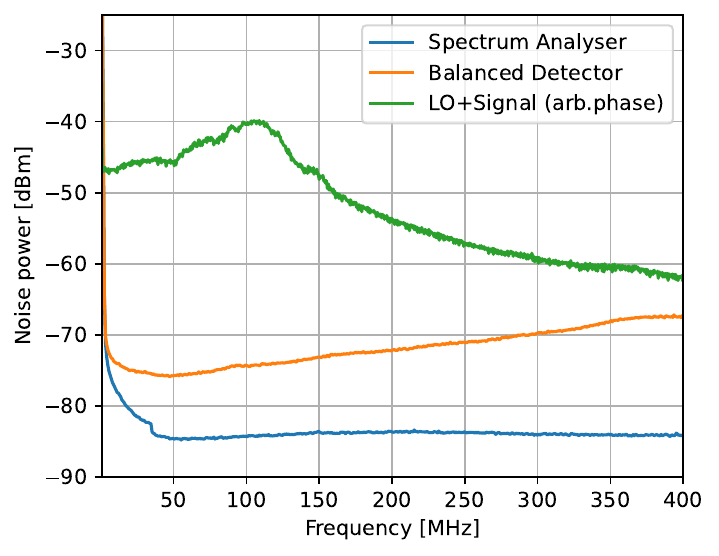}
	\caption{Spectra measured at the spectrum analyzer using only the spectrum analyzer (blue), the balanced detector (orange) and the LO mixed with a signal (green). }\label{fig:sa_noise}
\end{figure}

The experiments are conducted using a high-power pump laser source (Teraxion PureSpectrum LM-1550-12), which can be attenuated with a Thorlabs VOA50-APC single-mode in-line variable attenuator. Additionally, a low-power laser source (Agilent 81682A TLS in 8164B mainframe) is employed for the LO, with its phase adjustable via a phase shifter (General Photonics GPC-FPS-001). The out-coupled light from the chip is directed to a 90/10 beam splitter (Thorlabs TN1550R2A1), where 10 \% of the light is measured using a detector (Agilent 81619A Powermeter in an 8164B mainframe). The remaining 90 \% are then mixed at a 50/50 beam splitter (Thorlabs TN1550R5A1) and subsequently measured with a balanced detector (Thorlabs PDB470C). The whole experimental setup is connected via single-mode fibers (SMF-28) consisting of angled physical contacts (APC) to reduce the reflections inside of the system. The lensed fibers are positioned via mechanical stages (Scientific Elliot Gold Series XYZ Flexure Stages), which include piezoelectric actuators. This allows an efficient fiber-chip coupling. For all measurements it is necessary that the polarization inside of the chip is the same, since the resonator shows a polarization dependent behavior. Therefore, a polarization beam-splitter is integrated in the micro-chip which allows the adaption of the polarization inside the chip in combination with the polarization controller. All measurements are then performed using the transversal electric (TE) mode. \\
For the squeezing measurements, each measurement is conducted using the zero-span setting on a spectrum analyzer (Rhode \& Schwarz FSL3) at a frequency of 100 MHz, with a video bandwidth of 300 Hz and a resolution bandwidth of 300 kHz. To ensure accurate squeezing measurements, we first assess the noise of the system, as illustrated in Figure \ref{fig:sa_noise}. The results show that the signal mixed with the local oscillator (LO) is significantly above the system noise, which includes contributions from both the spectrum analyzer and the balanced detector.\\
The characterization measurements for generating the frequency comb are performed using a similar setup, but without the local oscillator (LO) and instead employing an optical spectrum analyzer (Advantest Q8384) in place of the balanced detector. In this case, the spectrum measurements are performed using the LO laser, which can be tuned over a wide wavelength range, rather than the pump laser. For transmission measurements at varying input powers, the LO laser is utilized in combination with an optical amplifier.

\subsection*{Dispersion engineering}

To analyze the optical nonlinearity $g_\mathrm{opt}$ of our system, we perform dispersion engineering to facilitate the generation of a frequency comb. The threshold power required to observe the first sidebands is influenced by the geometry, the pump frequency $\omega_{p}$ and $g_\mathrm{opt}$, as discussed in the main text. Therefore, generating a frequency comb is of particular interest. It is well established that anomalous dispersion is necessary in a ring resonator to enable the FWM process, and that the dispersion determines the spacing between the modes, expressed as $\omega_\mu = \omega_{0} + D_1\mu + 1/2D_2\mu^2+... = \omega_{0} + D_1\mu + D_\mathrm{int}$ \cite{FujiiTanabe+2020+1087+1104} with the mode number $\mu$. For our design, we utilize the mode solver FIMMWAVE to calculate the effective index $n_\mathrm{eff}$ across a range of wavelengths for various geometries. With the derived values of $n_\mathrm{eff}$, we can determine the dispersion. To validate our design, transmission measurements are conducted across multiple resonance frequencies, as shown in Figure \ref{fig:chip_charakterization}/a in the main text. The equidistant spacing between the modes yields a first-order dispersion of $D_1=0.68$ THz, while the higher-order dispersions are illustrated in the right plot of Figure \ref{fig:chip_charakterization}/b. The intrinsic dispersion $D_\mathrm{int}$ can be approximated by the second-order dispersion, resulting in $D_2 = 7.76$ MHz, which confirms the anomalous dispersion.  At this point, we stress again that we have decided to design the microring resonator in the anomalous dispersion regime only to measure the optical gain $g_\mathrm{opt}$ via the generation of a frequency comb. For any practical purposes, however, we recommend the normal dispersion regime as discussed above.

\subsection*{Setup losses}
The squeezing visibility is reduced due to losses and thus, it is critical to determine the efficiency $\eta$ from the squeezing generation to the detector. The influence on the generated squeezing up to the measurement is given as 
\begin{equation}
\langle V_\mathrm{s} \rangle_\mathrm{measured} = \left(1-\eta\right) + \eta \langle V_\mathrm{s} \rangle_\mathrm{Chip}.
\end{equation}
This is explained in more detail in the supplement materials and the losses lead to the requirement of a highly efficient setup. The losses of our system are characterized and listed in table \ref{table:losses}, which, in total results to an efficiency of $\eta = 28.46$\%. This efficiency is used to determine the squeezing inside the waveguide connected to the ring resonator to $-4.7$ dB
\begin{table}
	\begin{tabular}{|l|c|}
		\hline 
		\textbf{Loss source} & \textbf{Loss} \\ 
		\hline 
		Coupling from the waveguide to the fiber & -3.9 dB \\ 
		\hline 
		90/10 beam-splitter & -0.457 dB \\ 
		\hline 
		From the 90/10 beam-splitter to the detector & -1 dB \\ 
		\hline 
		\textbf{Total loss} &\textbf{ -5.357 dB}\\
		\hline 
	\end{tabular}\caption{Summary of the characterized losses of the measurement setup.}\label{table:losses}
\end{table}

\section*{Declarations}

\begin{footnotesize}

\noindent\textbf{Funding:}
The IPCEI ME/CT project is supported by the Federal Ministry for Economic Affairs and Climate Action on the basis of a decision by the German Parliament, by the Ministry for Economic Affairs, Labor and Tourism of Baden-W{\"u}rttemberg based on a decision of the State Parliament of Baden-W{\"u}rttemberg, the Free State of Saxony on the basis of the budget adopted by the Saxon State Parliament, the Bavarian State Ministry for Economic Affairs, Regional Development and Energy and financed by the European Union - NextGenerationEU. 

\noindent\textbf{Acknowledgments:}
We are very grateful for valuable discussions with Carlos Navarrete-Benlloch and Simon Abdani.

\noindent\textbf{Author contributions:} 
P.T. and P.DS. made the concept.
P.T, R.K. and C.S  designed the chip and the layout.
P.T and O.S. performed the measurements.
P.T, S.A. and P.DS analyzed the data.
P.T and P.DS wrote the manuscript.
P.DS, T.O., W.V, G.R. and A.Z. supervised all efforts.

\noindent\textbf{Competing interests:}
 Conflict of interest/Competing interests: The authors declare no competing interests.

\noindent\textbf{Data Availability Statement:} Data availability: The data sets generated during and/or analyzed during this study are available from the corresponding author on request.
\end{footnotesize}

\bibliography{chipSingleMode}
\bibliographystyle{apsrev4-2}
\end{document}


\title{ Supplementary Information: \it{Chip-integrated single-mode coherent-squeezed light source using four-wave mixing in microresonators}  }

\author{Patrick Tritschler}
\email{patrick.tritschler@de.bosch.com}
\affiliation{Robert Bosch GmbH, Robert-Bosch-Campus 1, Renningen, 71272, Germany}
\affiliation{Institute for Micro Integration (IFM), University of Stuttgart, Allmandring 9b, Stuttgart, 70569, Germany}

\author{Torsten Ohms}%
\affiliation{Bosch Sensortec GmbH, Gerhard-Kindler Stra{\ss}e 9, Reutlingen, 72770, Germany}

\author{Christian Schweikert}%
\affiliation{Institute of Electrical and Optical Communications, Pfaffenwaldring 47, 70569 Stuttgart, Germany}

\author{Onur S\"ozen}%
\affiliation{Institute of Electrical and Optical Communications, Pfaffenwaldring 47, 70569 Stuttgart, Germany}

\author{Rouven H. Klenk}%
\affiliation{Institute of Electrical and Optical Communications, Pfaffenwaldring 47, 70569 Stuttgart, Germany}

\author{Simon Abdani}%
\affiliation{Institute of Electrical and Optical Communications, Pfaffenwaldring 47, 70569 Stuttgart, Germany}

\author{Wolfgang Vogel}%
\affiliation{Institute of Electrical and Optical Communications, Pfaffenwaldring 47, 70569 Stuttgart, Germany}

\author{Georg Rademacher}%
\affiliation{Institute of Electrical and Optical Communications, Pfaffenwaldring 47, 70569 Stuttgart, Germany}

\author{Andr\'{e} Zimmermann}%
\affiliation{University of Stuttgart, Institute for Micro Integration (IFM) and Hahn-Schickard, Allmandring 9b, Stuttgart, 70569, Germany}

\author{Peter Degenfeld-Schonburg}%
\email{peter.degenfeld-schonburg@de.bosch.com}
\affiliation{Robert Bosch GmbH, Robert-Bosch-Campus 1, Renningen, 71272, Germany}

\maketitle

\renewcommand{\thefigure}{S\arabic{figure}}

\section*{Outra-cavity squeezing}\label{Ap1:OutraCavity}
To analyze the generation of squeezed light, we utilize the ring resonator model shown in Figure 1 in the main text. The resonator is pumped by the input mode $b_\mathrm{in} = \beta_\mathrm{in} + \delta b_\mathrm{in}$ with the coherent laser pump $\beta_\mathrm{in}=\sqrt{P_\mathrm{in}/\hbar\omega_{p}}$, where $P_\mathrm{in}$ represents the optical power and $\omega_{p}$ denotes the angular laser frequency utilized to excite the resonator mode $a_p$. This mode experiences losses characterized by $\gamma$ and $\kappa$. A portion of the mode $a_p$ is coupled out via $\kappa$ to produce the output light $P_\mathrm{out}$, in combination with the transmitted light. Within the ring, the four-wave mixing (FWM) process can occur. In this work, we focus on the process where two pump photons are absorbed and subsequently emitted, exploiting the third-order nonlinear susceptibility $\chi^{(3)}$, which is included into the optical gain $g_\mathrm{opt}$. According to \cite{PhysRevA.92.033840}, the system Hamiltonian for a single photonic mode within the cavity, characterized by the resonance frequency $\omega_p$, is expressed in the rotating wave approximation by
\begin{equation}
\begin{aligned}\label{equ:fwm_ham}
H_\mathrm{sys} = &\hbar\omega_R a_p^\dagger a_p - \frac{\hbar g_\mathrm{opt}}{2} a_p^\dagger a_p^\dagger a_p a_p
\end{aligned}
\end{equation}
with the resonance frequency of the ring resonator $\omega_R$. We emphasize that our description with a single-mode suffices for resonators in the normal dispersion regime where the excitation through FWM processes is strongly suppressed \cite{Herr2012}, especially in the case of our interest where the resonator is driven at the injection locking point. Based on this Hamiltonian and following \cite{Lambropoulos2007, Gardiner85, Collet84}, the quantum Langevin equations can be derived in the rotating frame approximation to
\begin{equation}\label{equ:quantumLangevin_SpmSqueezing}
\begin{aligned}
\frac{d}{dt}  a_{p}(t) = \left(\omega_p -\omega_{R} + g_{th}\langle a_p^\dagger a_p\rangle + g_{\mathrm{opt}} a_p^\dagger a_p \right) i a_p - \frac{\Gamma}{2} a_p + \sqrt{\kappa} b_{in}(t) + \sqrt{\gamma} b_{\gamma}(t)
\end{aligned}
\end{equation}
with the total loss $\Gamma=\kappa +\gamma$ and the vacuum mode $b_\gamma$ that couples through $\gamma$. Following \cite{tritschler2024nonlinearopticalbistabilitymicroring}, an additional detuning term caused by thermal self-heating scaling with the thermal gain $g_\mathrm{th}$ is added to equation \ref{equ:quantumLangevin_SpmSqueezing}. As commonly done, we separate the bosonic light field $a_p = \alpha_p + \delta a_p$ into its coherent part $\alpha_p$ and its quantum fluctuations $\delta a_p$. Next, we linearize the the Langevin equations around the fluctuations $\delta a_p$ of the field, which for example gives
\begin{equation}\label{linearizationExample}
a_p^\dagger a_p a_p = ( \delta a_p^\dagger + \alpha_{p}^* )( \delta a_p + \alpha_{p} )( \delta a_p + \alpha_{p} ) \approx \delta a_p^\dagger \alpha_{p}^2 + 2 \alpha_{p}^* \alpha_{p} \delta a_p + \alpha_p^2 \alpha_{p}^*
\end{equation}
with neglecting higher order terms of the fluctuations. This linearization is also performed for the input and the vauum mode with $b_\mathrm{in} = \beta_\mathrm{in} + \delta b_\mathrm{in}$ and $b_\gamma = \beta_\gamma + \delta b_\gamma$, which leads to the following quantum Langevin equations for the coherent part and the fluctuations with
\begin{eqnarray}
\frac{d }{dt}\alpha_{p} & = & i \Delta_\mathrm{cl} \alpha_{p} - \frac{\Gamma}{2} \alpha_p + \sqrt{\kappa} \beta_{in} + \sqrt{\gamma} \beta_{\gamma}, \label{equ:quantumLangevin_SpmClassic_linearized}\\
\frac{d }{dt} \delta a_{p} & = & i\Delta_f  \delta a_p + ig_{\mathrm{opt}} \delta a_p^\dagger \alpha_{p}^2-  \frac{\Gamma}{2} a_p + \sqrt{\kappa} \delta b_{in} + \sqrt{\gamma}\delta b_{\gamma}, \label{equ:quantumLangevin_SpmSqueezing_linearized}
\end{eqnarray}
with the classical detuning 
\begin{equation}
\Delta_\mathrm{cl} =  \omega_{p}-\omega_{R} + |\alpha_p|^2 (g_\mathrm{opt} + g_\mathrm{th})
\end{equation}
and the detuning for the fluctuations
\begin{equation}
\Delta_f = \Delta_\mathrm{cl} + g_{\mathrm{opt}} |\alpha_{p}|^2.
\end{equation}
The key distinction between the equations of motion for the coherent part and the fluctuations lies in the generation of the fluctuations, as well as the additional detuning of the fluctuations, which both depends on $g_\mathrm{opt}$. Both factors emerge during the linearization process described in equation \ref{linearizationExample} and are crucial for the subsequent analysis of squeezing. Particularly significant is the additional detuning given by $g_\mathrm{opt}|\alpha_p|^2$. Even at the injection locking point where $\Delta_\mathrm{cl} = 0$, the fluctuations become detuned. We will analyze the impact of this detuning on the quality of squeezing in the following.\\
At the injection locking point we have $\Delta_\mathrm{cl} = 0$ and from equation \ref{equ:quantumLangevin_SpmClassic_linearized}, we immediately find the steady state of the classical field amplitude $\alpha_p = 2\sqrt{\kappa}\beta_\mathrm{in}/\Gamma$. Moreover, it is then straight forward to solve equation \ref{equ:quantumLangevin_SpmSqueezing_linearized} for the second order moments of the fluctuations. The photon number of the fluctuations is for example given by $\langle\delta a_p ^\dagger \delta a_p\rangle = 1/2 \cdot |x|^2 / (1-|x|^2)$ with the distance to the threshold $0 < x < 1$. It can be shown that $x$ at the injection locking point is given in physical units by $|x|^2 = 1/(1+(P_\mathrm{th}/P_\mathrm{in})^2)$ with $P_\mathrm{th} = \Gamma^3 \hbar \omega_p /(8g_\mathrm{opt}\kappa)$. The quantity $P_\mathrm{th}$ would correspond to the threshold of multi-mode FWM in the case of a resonator with anomalous dispersion. Clearly, we see that $x<1$ for all input powers. Therefore, our system in the normal dispersion regime and at the injection locking point is always stable and the classical part of the output light, and thus, the brightness can be controlled well by the input laser power. As an additional remark, we would like to state that the validity of the linearized Quantum Langevin equations has its limit. It will brake down very close to the critical point \cite{tritschler2024optical, PhysRevA.93.023819, PhysRevA.91.053850}, which we expect at $x \approx 0.995$. This, however, corresponds to an input power of $P_\mathrm{in} \approx 80mW$, clearly highlighting both the validity of the approach we use in this work and the robustness of our system. We proceed with the fluctuations using the input-output theory with $b_{\mathrm{out}} = \sqrt{\kappa} a_{p} - b_{\mathrm{in}}$ \cite{Collet84, Gardiner85}, to describe the cavity mode in dependency of the input and the output mode with
\begin{eqnarray}
\frac{d}{dt} \mathbf{A}(t) &=& \left(\mathbf{T}-\frac{\kappa}{2} \mathbf{1}\right) \mathbf{A}(t) +  \sqrt{\kappa} \mathbf{B_{in}}(t) + \sqrt{\gamma} \mathbf{B_{\gamma}}(t), \label{equ:eqm_t_in_sm}\\ 
\frac{d}{dt} \mathbf{A}(t) &=& \left( \mathbf{T}+\frac{\kappa}{2} \mathbf{1}\right) \mathbf{A}(t) - \sqrt{\kappa} \mathbf{B_{out}}(t) + \sqrt{\gamma} \mathbf{B_{\gamma}}(t)\label{equ:eqm_t_out_sm}
\end{eqnarray}
and with the transfer matrix
\begin{equation}\label{equ:transferMatrix_sm}
\begin{aligned}
T = \left(\begin{matrix}
-i\left( \Delta_\mathrm{cl} + g_{\mathrm{opt}} |\alpha_{p}|^2 \right) - \frac{\gamma}{2}  & -i\frac{\sigma}{2}  \\
i\frac{\sigma}{2}  & i\left( \Delta_\mathrm{cl} + g_{\mathrm{opt}} |\alpha_{p}|^2 \right) - \frac{\gamma}{2},
\end{matrix}
\right),
\end{aligned}
\end{equation}
the system vectors
\begin{eqnarray}
\mathbf{A} = \left(\begin{array}{c}
\delta a_p\\
\delta a_p^{\dagger} \\
\end{array}\right), \quad
\mathbf{B_{in}} = \left(\begin{array}{c}
\delta b_{in}\\
\delta b_{in}^{\dagger} \\
\end{array}\right), \quad
\mathbf{B_{\gamma}} = \left(\begin{array}{c}
\delta b_{\gamma}\\
\delta b_{\gamma}^{\dagger} \\
\end{array}\right), \quad
\mathbf{B_{out}} = \left(\begin{array}{c}
\delta b_{out}\\
\delta b_{out}^{\dagger} \\
\end{array}\right)
\end{eqnarray}
and the injection parameter $\sigma = 2g_{opt} \cdot \alpha_p^2$. The equations \ref{equ:eqm_t_in_sm} and \ref{equ:eqm_t_out_sm} can be solved in the frequency space and rearranged to eliminate $\mathbf{A}$, which leads to the following description of the output mode
\begin{eqnarray}\label{equ:outputMode_sm}
\begin{aligned}
\mathbf{B_{out}}(\omega) =- \frac{1}{\sqrt{\kappa}}  \Bigl\lbrack[\mathbf{\Omega}-\mathbf{T}-\frac{\kappa}{2} \mathbf{I_2}][\mathbf{\Omega}-\mathbf{T}+\frac{\kappa}{2} \mathbf{I_2}]^{-1} \\
\cdot\left( \sqrt{\kappa}\mathbf{B_{in}}(\omega) + \sqrt{\gamma} \mathbf{B_{\gamma}}(\omega) \right)- \sqrt{\gamma} \mathbf{B_{\gamma}}(\omega) \Bigr\rbrack
\end{aligned}
\end{eqnarray}
with the frequency vector
\begin{eqnarray}
\mathbf{\Omega} = \left(\begin{array}{c}
\omega\\
\omega \\
\end{array}\right).
\end{eqnarray} 
Equation \ref{equ:outputMode_sm} describes the output modes and includes the losses within the ring resonator. However, once these modes couple from the resonator to the waveguide to form the mode $b_\mathrm{out}$, they are subject to additional losses before being detected. This can be modeled using a beam splitter that combines the output modes with vacuum noise, represented as $\sqrt{\eta}\mathbf{B_{out}}(\omega) + \sqrt{1-\eta}\mathbf{B_v}$, where the vacuum vector is defined as $\mathbf{B_v}=[b_v, b_v^\dagger]^\top$ and $\eta$ denotes the efficiency. By applying the canonical commutation relations of the vacuum modes and following the methodologies outlined in \cite{zoller1997quantum, wiseman_milburn_2009}, we can determine the expectation values of the vacuum modes as
\begin{eqnarray} \label{equ:vacuum_exp}
\begin{aligned}
\langle b_v( t) \rangle &=  \langle b_v^{\dagger}(t) \rangle = \langle b_v( t) b_v( t' ) \rangle = \langle b_v^{\dagger}( t) b_v( t' ) \rangle = 0& \\ \langle b_v\left( t \right) b_v^{\dagger}( t' ) \rangle &= \delta(t-t').&
\end{aligned}
\end{eqnarray}
Since only vacuum modes appear in equation \ref{equ:outputMode_sm}, all expectation values of the outra-cavity modes can be determined with equation \ref{equ:vacuum_exp}. Thus, we can solve equation \ref{equ:outputMode_sm} for example for the number of fluctuating photons to
\begin{equation}\label{equ:numberFluctPhotons}
\langle\delta b_\mathrm{out}^\dagger \delta b_\mathrm{out}\rangle = \frac{4\eta\kappa\sigma^2\Gamma}{ \left(4 \Delta_f^2 +\Gamma^2 - \sigma^2 \right)^2  } \delta(\omega-\omega').
\end{equation}
For single-mode squeezed light, the following definition of the variance can be used to analyze the squeezing with \cite{Walls2008}
\begin{equation}\label{equ:variance}
\begin{aligned}
V\left(\omega,\omega'\right) = \Delta X_Q\left(\omega,\omega'\right) = \langle X_Q\left(\omega\right) X_Q\left(\omega'\right) \rangle - \langle X_Q\left(\omega\right) \rangle\langle X_Q\left(\omega'\right) \rangle 
\end{aligned}
\end{equation}
and the quadrature operator 
\begin{equation} \label{equ:quadrature_signal_sm}
X_Q(\omega) = \frac{1}{\sqrt{2}}\left[ \delta b_{out}  e^{i \phi_{LO}} + \delta b_{out}^{\dagger} e^{-i \phi_{LO}}\right]. 
\end{equation}
We solve equation \ref{equ:variance} by introducing the dimensionless injection parameter
\begin{equation}
\tilde{\sigma} = \frac{\sigma}{\Gamma},
\end{equation}
the dimensionless frequency
\begin{equation}
y = 1 + \left( \frac{2\omega}{\Gamma}\right)^2
\end{equation}
and the dimensionless coefficient 
\begin{equation}
c = \frac{4\eta\kappa}{\Gamma},
\end{equation}
which leads to the following expression for the single-mode squeezing variance normalized to the vacuum variance and at $\omega=0$ with
\begin{equation}\label{equ:squeezingRawEq}
\frac{\langle V \rangle }{\langle V_\mathrm{vac}\rangle} = 1 + \frac{c}{y^2}\left( \left[ \frac{i\tilde{\sigma}}{2} \left(y^2+2i\tilde{\sigma}\right) \right] e^{-2i\phi_{LO}} - \left[ \frac{i\tilde{\sigma}}{2} \left(y^2-2i\tilde{\sigma}\right) \right] e^{2i\phi_{LO}} + 2\tilde{\sigma}^2 \right).
\end{equation}
The full solution of the variance $V$ with the equation \ref{equ:squeezingRawEq} is shown in the Figure 2/B in the main text in dependency of the detuning of the pump frequency to the cold cavity frequency $\Delta_p=\omega_{p}-\omega_{R}$ and the squeezing angle $\phi_{LO}$ with $\kappa=500$ MHz, $\gamma=50$ MHz, $g_\mathrm{opt}=1.5$ Hz and $g_\mathrm{th}=100$ Hz. Thereby, a scanning of the resonance frequency with a lowered frequency is assumed and the best squeezing is achieved at the injection locking point $\Delta_\mathrm{cl}=0$ with a broadband spectrum that can be observed. It is important to note, that the angle $\phi_{LO}$ to achieve the best squeezing depends on the resonator geometry and the input power with  
\begin{equation}\label{equ:optimalPhase}
\phi_{\mathrm{LO, opt}} = \frac{1}{2}\tan^{-1}\left(-\frac{1}{2\tilde{\sigma}} \right).
\end{equation}
The angle for the anti-squeezing is shifted by $\pi/2$ with respect to $\phi_{\mathrm{LO, opt}}$. Remarkably, we can drastically simplify the squeezing and anti-squeezing at the injection locking point to give the result as already stated in the main text in equations 3 and 4. Finally, we give the number of the output photon fluctuations at the injection locking point by 
\begin{equation}
\langle\delta b_\mathrm{out}^\dagger \delta b_\mathrm{out}\rangle = \frac{4\eta\kappa}{\Gamma} \left(\frac{P_\mathrm{in}}{P_\mathrm{th}}\right)
\end{equation}
which also gives the squeezing parameter by a comparison with \cite{Gerry_Knight_2004} to $r = \sinh^{-1}\left( \langle\delta b_\mathrm{out}^\dagger \delta b_\mathrm{out}\rangle \right)$. 
\begin{figure}
	\centering
	\includegraphics[width=0.001\textwidth]{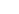}
\end{figure}

\bibliography{chipSingleMode} 
\bibliographystyle{naturemag}